\documentstyle[12pt]{article}

\catcode`\@=11

\@addtoreset{equation}{section}
\def\theequation{\arabic{section}.\arabic{equation}}

\def\@normalsize{\@setsize\normalsize{15pt}\xiipt\@xiipt
\abovedisplayskip 14pt plus3pt minus3pt%
\belowdisplayskip \abovedisplayskip
\abovedisplayshortskip  \z@ plus3pt%
\belowdisplayshortskip  7pt plus3.5pt minus0pt}

\def\small{\@setsize\small{13.6pt}\xipt\@xipt
\abovedisplayskip 13pt plus3pt minus3pt%
\belowdisplayskip \abovedisplayskip
\abovedisplayshortskip  \z@ plus3pt%
\belowdisplayshortskip  7pt plus3.5pt minus0pt
\def\@listi{\parsep 4.5pt plus 2pt minus 1pt
            \itemsep \parsep
            \topsep 9pt plus 3pt minus 3pt}}

\def\underline#1{\relax\ifmmode\@@underline#1\else
        $\@@underline{\hbox{#1}}$\relax\fi}
\@twosidetrue

\relax

\catcode`@=12

\catcode`\@=11

\def\section{\@startsection{section}{1}{\z@}{3.5ex plus 1ex minus
   .2ex}{2.3ex plus .2ex}{\large\bf}}

\def\thesection{\Roman{section}.}

\def\appendix{\setcounter{section}{0}
        \def\thesection{APPENDIX }
        \def\theequation{\Alph{section}.\arabic{equation}}}

\def\FERMIPUB{}

\def\ps@headings{\def\@oddfoot{}\def\@evenfoot{}
\def\@oddhead{\hbox{}\hfill
        \makebox[.5\textwidth]{\raggedright\ignorespaces --\thepage{}--
        \hfill {\rm FERMILAB--Pub--\FERMIPUB}}}
\def\@evenhead{\@oddhead}
\def\subsectionmark##1{\markboth{##1}{}}
}

\catcode`\@=12

\relax

\def\figcap{\section*{Figure Captions\markboth
        {FIGURECAPTIONS}{FIGURECAPTIONS}}\list
        {Fig. \arabic{enumi}:\hfill}{\settowidth\labelwidth{Fig. 999:}
        \leftmargin\labelwidth
        \advance\leftmargin\labelsep\usecounter{enumi}}}
 \relax
\def\tablecap{\section*{Table Captions\markboth
        {TABLECAPTIONS}{TABLECAPTIONS}}\list
        {Table \arabic{enumi}:\hfill}{\settowidth\labelwidth{Table 999:}
        \leftmargin\labelwidth
        \advance\leftmargin\labelsep\usecounter{enumi}}}
 \relax
\def\reflist{\section*{References\markboth
        {REFLIST}{REFLIST}}\list
        {[\arabic{enumi}]\hfill}{\settowidth\labelwidth{[999]}
        \leftmargin\labelwidth
        \advance\leftmargin\labelsep\usecounter{enumi}}}
 \relax

\catcode`\@=11

\def\FERMIPUB{}

\def\ps@headings{\def\@oddfoot{}\def\@evenfoot{}
\def\@oddhead{\hbox{}\hfill
        \makebox[.5\textwidth]{\raggedright\ignorespaces --\thepage{}--
        \hfill {\rm FERMILAB--Pub--\FERMIPUB}}}
\def\@evenhead{\@oddhead}
\def\subsectionmark##1{\markboth{##1}{}}
}

\ps@headings

\relax
\newskip\humongous \humongous=0pt plus 1000pt minus 1000pt
\def\caja{\mathsurround=0pt}
\def\eqalign#1{\,\vcenter{\openup1\jot \caja
        \ialign{\strut \hfil$\displaystyle{##}$&$
        \displaystyle{{}##}$\hfil\crcr#1\crcr}}\,}
\newif\ifdtup

\def\beq{\begin{equation}}
\def\eeq{\end{equation}}

\def\beqn{\begin{eqnarray}}
\def\eeqn{\end{eqnarray}}
\relax

\def\G2{{\; \rm GeV/}c^2}
\def\G{\; \rm GeV}

\def\dotx{\dotx{\dot\overline{x}}}

\relax

\begin{document}
\hbadness=10000
\begin{titlepage}
\nopagebreak
\begin{flushright}

        {\normalsize KUCP-47\\

        May,~1992}\\
\end{flushright}
\vfill
\begin{center}
{\large \bf Differential Calculus on the Quantum \\ Superspace and
Deformation of Phase Space}
\vfill
\renewcommand{\thefootnote}{\fnsymbol{footnote}}
{\bf Tatsuo Kobayashi\footnote{
Fellow of the Japan Society for the Promotion of Sience. Work
partially supported by the Grant-in-Aid for Scientific Research from the
Ministry of Education, Science and Culture (\# 030083)} and Tsuneo Uematsu
\footnote{Work partially supported by the Grant-in-Aid for Scientific Research
from the Ministry of Education, Science and Culture (\# 04245221)}}

       Department of Physics, College of Liberal Arts and Sciences, \\
       Kyoto University,~Kyoto 606,~Japan \\
\vfill

\end{center}

\vfill
\nopagebreak
\begin{abstract}
We investigate non-commutative differential calculus on the supersymmetric
version of quantum space in which quatum supergroups are realized.
Multi-parametric quantum deformation of the general linear supergroup,
$GL_q(m|n)$, is studied and the explicit form for the ${\hat R}$-matrix is
presented.  We apply these results to the quantum phase-space construction of
$OSp_q(2n|2m)$ and calculate their ${\hat R}$-matrices.

\end{abstract}

\vfill
\end{titlepage}
\pagestyle{plain}
\newpage
\voffset = -2.5 cm
\leftline{\large \bf 1. Introduction}
\vspace{0.8 cm}

  Recently there has been much interest in quantum groups
\cite{Drinfeld,Jimbo,Faddeev-Reshetikhin-Takhtajan}, in the context of
studying integrable field theories, statistical models and conformal
field theories in two dimensions (See, e.g. refs. \cite{Doebner-Hennig,CFZ}).

  Quantum groups can be realized on  the quantum (hyper-)plane, in which
coordinates are non-commuting \cite{Manin,Takhtajan}.
Woronowicz has initiated the differential calculus on the non-commutative
space of quantum groups \cite{Woronowicz}, which provides an example for
non-commutative differential geometry \cite{Connes}.
  Wess and Zumino and others developed the differential calculus on the
quantum (hyper-)plane covariant with respect to quantum groups
\cite{Wess-Zumino} and multiparameter deformation of the quantum groups,
especially for $GL_q(n)$
\cite{Schirrmacher-Wess-Zumino,Schirrmacher1,Schirrmacher2},
as well as for $SO_q(n)$ \cite{Carow-Watamura-Schlieker-Watamura}.

  Now it would be interesting to study q-deformation of supergroups from the
viewpoint of quantum space.  Quantum Lie superalgebras were studied in the
framework of bosonic and fermionic q-oscillators
\cite{Kulish,Chaichan-Kulish,Chaichan-Kulish-Lukierski}.
The Manin's construction of quantum groups has already been extended
to supergroups without recourse to differential calculus
\cite{Corrigan-Fairlie-Fletcher-Sasaki}.

 In this paper we investigate q-deformation of supersymmetric groups in
the framework of differential calculus on the supersymmetric version of
quantum hyper-plane or quantum superspace, where the non-commuting
super-coordinates consist of bosonic as well as fermionic (Grassmann)
coordinates. Based on the non-commutative differential calculus on the
quantum superspace, we study quantum deformation of supergroups and
${\hat R}$-matrices which are solutions of the Yang-Baxter equation. We shall
obtain the explicit form of the multiparametric ${\hat R}$-matrices for the
quantum supergroup $GL_q(m|n)$.  In contrast to the Manin's superspace
\cite{Manin,Corrigan-Fairlie-Fletcher-Sasaki} we have the
differential structure of the quantum superspace, therefore we can extend our
results to quantum phase space of coordinates and momenta
\cite{Wess-Zumino,Zumino1,Aref'eva-Volovich}.
Actually we shall apply our results for the $GL_q(m|n)$ to the phase space
construction \cite{Zumino1} of $OSp(2n|2m)$-type quantum supergroups.
\cite{Kulish2,Saleur,Deguchi-Fujii-Ito}.

 This paper is organized as follows.  In the next section we introduce the
quantum superspace and study the differential calculus on it, in connection
with the Yang-Baxter equation. In section 3, the commutation relation of
the matrix elements of quantum supergroups together with the quantum
superdeterminant are investigated. Based on these results we construct
the quantum phase-space of the $GL_q(m|n)$ in section 4.  The final section
is devoted to some concluding remarks.

\vspace{0.8 cm}
\leftline{\large \bf 2. Quantum superspace}
\vspace{0.8 cm}

Now let us introduce the supersymmetric version of non-commuting
coordinates which characterize the quantum supersymmetric hyper-plane
or quantum superspace as follows:
$$
Z^{I} = (x^{i}; \theta^{\alpha}) \qquad  (i=1, \cdots , m \ ; \alpha =
1, \cdots , n)
\eqno(2.1)$$
where $x^{i}$ and $\theta^{\alpha}$ denote bosonic and fermionic (Grassmannian)
coordinates, respectively. They satisfy the following commutation relation
$$
Z^{I} Z^{J} = B^{IJ}_{\ \ KL} Z^{K} Z^{L}.
\eqno(2.2)$$
We also set up the commutation relation between coordinates and differentials:
$$
Z^{I} dZ^{J} = C^{IJ}_{\ \ KL} dZ^{K} Z^{L}.
\eqno(2.3)$$
We will see that these two coefficients $B^{IJ}_{\ \ KL}$ and $C^{IJ}_{\ \ KL}$
are mutually related with each other from the consistency conditions.
We now extend the differential calculus on quantum space developed by Wess and
Zumino \cite{Wess-Zumino} and others [11-14] to this quantum superspace.
Here we require that the exterior derivative $d$ given by
$$
d=dZ^{I}{\partial\over {\partial Z^I}}=dx^i{\partial\over {\partial x^i}}
+ d\theta^{\alpha}{\partial\over {\partial \theta^{\alpha}}}
\eqno(2.4)$$
obey the following two conditions:

i) nilpotency : $d^2=0$

ii) Leibniz rule: $d(fg) = (df)g + (-1)^{{\hat f}}f(dg)$
where ${\hat f}$ denotes the Grassmann parity i.e. if $f$ is bosonic ${\hat f}
=0 $ and ${\hat f}=1 $ if $f$ is a Grassmann variable.

{}From the Leibniz rule we have for an arbitrary function $f$
$$d(Z^I f) = dZ^I({\partial\over {\partial Z^J}}Z^I)f +(-1)^{\hat I}Z^IdZ^J
{\partial\over {\partial Z^J}}f.
\eqno(2.5)$$
Hence by using the commutation relation between the coordinates and the
differentials (2.3) we get
$${\partial\over {\partial Z^J}}Z^I=\delta^I_{\ J} +
(-1)^{\hat I}C^{IK}_{\ \ JL}Z^L{\partial\over {\partial Z^K}}.
\eqno(2.6)$$
We find the following equation holds
$$
{\partial\over {\partial Z^R}}dZ^K = D^{KM}_{\ \ \ RN}dZ^N
{\partial\over {\partial Z^M}}
\eqno(2.7)$$
where
$$
D^{KM}_{\ \ \ RN} = (C^{-1})^{KM}_{\ \ \ RN}.
\eqno(2.8)$$
This relation can be obtained by multiplying the both sides of (2.7)
by $Z^M$ from the right and by making use of (2.6) together with the
inverse of (2.3).
Moreover, we multiply (2.3) by $\partial/{\partial Z^R}$ from the
left and commute this derivative through to the right by using (2.6) and (2.7)
we obtain
$$
(-1)^{\hat I}C^{IM}_{\ \ RM}D^{JS}_{\ \ \ MT}C^{NT}_{\ \ UV}
= (-1)^{\hat L}C^{IJ}_{\ \ KL}D^{KM}_{\ \ \ RU}C^{LS}_{\ \ MV}.
\eqno(2.9)$$
{}From the relation (2.8) we get
$$
(-1)^{\hat I}C^{IJ}_{\ \ KL}C^{LN}_{\ \ SV}C^{KS}_{\ \ MQ}
= (-1)^{\hat T}C^{JN}_{\ \ SU}C^{IS}_{\ \ MT}C^{TU}_{\ \ QV}.
\eqno(2.10)$$

Now we consider the relation between $B^{IJ}_{\ \ KL}$ and $C^{IJ}_{\ \ KL}$.
Acting a partial derivative $\partial/ {\partial Z^M}$ on the
commutation relation (2.2) from the left we extract the relation
$$
(\delta^I_{\ K}\delta^J_{\ L} - B^{IJ}_{\ \ KL})(\delta^K_{\ M}\delta^L_{\ T}
+ (-1)^{\hat K}C^{KL}_{\ \ MT}) = 0.
\eqno(2.11)$$
We further get a relation by taking the partial derivative of the relation
(2.2) multiplied by an arbitrary function of $Z$ and using the commutaion
relation (2.2) as
$$
(-1)^{{\hat I}+{\hat J}}C^{IP}_{\ \ MT}C^{JN}_{\ \ PU}B^{TU}_{\ \ QV}
= (-1)^{{\hat K}+{\hat L}}B^{IJ}_{\ \ KL}C^{KS}_{\ \ MQ}C^{LN}_{\ \ SV}.
\eqno(2.12)$$
Now if we take
$$
B^{IJ}_{\ \ KL} = (-1)^{\hat I}{1\over X}C^{IJ}_{\ \ KL},
\eqno(2.13)$$
where $X$ is an arbitrary parameter, and substitute it into (2.12) we can
recover the relation (2.10).  Here we should note that the conservation of
Grassmann parity : ${\hat I}+{\hat J}={\hat K}+{\hat L}$ holds for the
relation (2.2) as well as for the above equation. This can be proved
from another derivation of the same relation by multiplying the differential
of $Z$ on (2.2) from the right and commuting it through to the left.
The parameter $X$ will turn out to be one of the deformation parameters.
Thus we are led to the Yang-Baxter equation which is satisfied by
$B^{IJ}_{\ \ KL}$ given by (2.13)
$$
B^{IJ}_{\ \ KL}B^{LN}_{\ \ SV}B^{KS}_{\ \ MQ}
= B^{JN}_{\ \ PU}B^{IP}_{\ \ MT}B^{TU}_{\ \ QV},
\eqno(2.14)$$
which can be illustrated in Fig.1.
This equation can be cast in the more familiar Yang-Baxter equation
in terms of ${\hat R}$-matrices which are defined with
the conventional normalization as follows
$$
B^{IJ}_{\ \ KL}={1\over q}{{\hat R}^{IJ}}_{\ \ KL}
\quad, \quad q_{IJ} \equiv q.
\eqno(2.15)$$
In a similar argument which leads to eqs.(2.11) and (2.13), we can derive
the following commutation relation:
$$
\partial_I\partial_J = B^{LK}_{\ \ JI}\partial_K \partial_L \quad,
\quad \partial_I \equiv \partial/\partial Z^I.
\eqno(2.16)$$

Now let us restrict ourselves to the case of the q-deformed general linear
supergroup $GL_q(m|n)$, for which we set up the following commutation
relations
$$\eqalign{
Z^IZ^J &= q_{IJ}Z^JZ^I \cr
dZ^IdZ^J &=-p_{IJ}dZ^JdZ^I. \cr
}\eqno(2.17)$$
Taking the exterior derivative of (2.3), generally we have
$$
dZ^IdZ^J = -(-1)^{\hat K}C^{IJ}_{\ \ KL}dZ^KdZ^L.
\eqno(2.18)$$
Noting that $(dx^i)^2=0$, $(\theta^{\alpha})^2=0$ and using the relation
(2.13) we explicit worked out the computation of $C^{IJ}_{\ \ KL}$.
It turns out that $q_{IJ}$ and $p_{IJ}$ are related
with each other by the equation
$$p_{IJ} =(-1)^{{\hat I}+{\hat J}}{1\over X}q_{IJ},
\eqno(2.19)$$
and the explict form for the $B^{IJ}_{\ \ KL}$ or ${\hat R}$-matrix is
found to be
$$
B^{IJ}_{\ \ KL} = {1\over q}{{\hat R}^{IJ}}_{\ \ KL}
=\delta^I_{\ L}\delta^J_{\ K} \Bigl ( \bigl (-{1\over X} \bigr )^{\hat I}
\delta^{IJ} + \Theta^{IJ}{1\over q_{JI}}+\Theta^{JI}{q_{IJ}\over X} \Bigl )
+\delta^I_{\ K}\delta^J_{\ L}\Theta^{JI} \Bigl ( 1-{1\over X} \Bigl )
\eqno(2.20)$$
where $\Theta^{IJ}$ is equal to 1 for $I>J$ and 0 for $I \leq J$, and $q$
denotes $q_{IJ}$.
This solution for the Yang-Baxter equation provides the multiparametric
deformation of $GL(m|n)$ with $q_{IJ}$ and $X$ as the deformation parameters.
In the limit in which $q_{IJ}$ and $X$ tend to 1 in (2.20), we can recover
the classical case.
Note that the whole $B$-matrix is composed of submatrices for a pair of the
indices $I$ and $J$ ($I < J$) and is given by
$$
 B_{(I,J)} =
\left(\begin{array}{cccc}
\bigl (-{1\over X}\bigr )^{\hat I} & 0 & 0 & 0 \\
0 & 1-{1\over X} & {q\over X} & 0 \\
0 & {1\over q} &0 & 0  \\
0 & 0 & 0 & \bigl (-{1\over X}\bigr )^{\hat J}
\end{array}\right)
\eqno(2.21)$$
and we find the eigenvalues for this submatrix are 1, $-1/X$,
$(-1/X)^{\hat I}$ and $(-1/X)^{\hat J}$.
Therefore, the eigenvalue equation of the whole $B$-matrix for $GL_q(m|n)$
($m+n=N$) is given by
$$
\{(\lambda -1)(\lambda+{1\over X})\}^{N(N-1)/2}
(\lambda -1)^m(\lambda+{1\over X})^n = 0.
\eqno(2.22)$$
Hence the characteristic equation satisfied by the ${\hat R}$-matrix is seen
to be
$$
(\delta^I_{\ K}\delta^J_{\ L} -B^{IJ}_{\ \ KL})
(\delta^K_{\ M}\delta^L_{\ T}+XB^{KL}_{\ \ MT})=0
\eqno(2.23)$$
which is again consistent with (2.11) and (2.13).  Note that for the
case of $GL(2|0)$ ($GL(n|0)$), (2.20) reduces to the well-known result
for ${\hat R}$-matrix of the 2-parameter (the multiparameter)
deformation of $GL(2)$ \cite{Schirrmacher-Wess-Zumino} ($GL(n)$
\cite{Schirrmacher1,Schirrmacher2}).(The multiparameter deformation of
$GL(n)$ was also considered in ref.\cite{Fairlie-Zachos} on the basis of
q-oscillators.)
For $GL(2)$, the choice :$X=q^2$
leads to the well-known ${\hat R}$-matrix with one deformation parameter
$q$ \cite{Wess-Zumino}.  One-parameter deformed $GL(1|1)$, also for the
case $X=q^2$, was discussed in ref. \cite{Schmidke-Vokos-Zumino}.
By introducing projection operators ${\cal S}$ and ${\cal A}$ given as
$$
{\cal S}={1\over {1+1/X}}(B+{1\over X}{\bf 1}) \quad,\quad
{\cal A}={{-1}\over {1+1/X}}(B-{\bf 1}),
\eqno(2.24)$$
with the properties ${\cal S}^2={\cal S}$, ${\cal A}^2={\cal A}$,
${\cal S}+{\cal A}={\bf 1}$ and ${\cal A}{\cal S}={\cal S}{\cal A}=0$,
the $B$-matrix can be expressed as
$$
B = {\cal S}-{1\over X}{\cal A}.
\eqno(2.25)$$

  Now we make some remarks on ordering of the indices.
We distinguish two types of ordering between bosonic and fermionic(Grassmann)
coordinates for ${\hat R}$-matrices of $GL_q(m|n)$ ,
Type I : $i < \alpha $ and Type II : $ i > \alpha$.  Although this is not
essential for the quantum-matrix commutation relation,
these types lead to different expressions of the commutation relation
between coordinates and partial derivatives given in (2.6).

Choosing the type I ordering, we can explicitly write down the expression
of (2.6) for $GL_q(m|n)$ as follows
$$\eqalign{
& \partial_i x^i = 1 + X x^i\partial_i + (X-1)\sum_{a=i+1}^m x^a\partial_a
+ (X-1)\sum_{\alpha=1}^n \theta^{\alpha}\partial_{\alpha} \cr
& \partial_JZ^I=q_{IJ}Z^I\partial_J \quad , \quad
\partial_IZ^J={X\over q_{IJ}}Z^J\partial_I \quad , \quad I<J\cr
& \partial_{\alpha}\theta^{\alpha} = 1 - \theta^{\alpha}\partial_{\alpha}
-(1-X)\sum_{\beta=\alpha +1}^n \theta^{\beta}\partial_{\beta}. \cr
}\eqno(2.26)$$
A similar but different expression can also be obtained for the type II
ordering.

\vspace{0.8 cm}
\leftline{\large \bf 3. Quantum group and superdeterminant}
\vspace{0.8 cm}

The coordinates of the superspace, $Z^J$ are linearly transformed by a matrix
$T^I_{\ J}$ into new coordinates $Z'^I$, i.e., $Z'^I=T^I_{\ J}Z^J$.
Covariance of (2.2) requires commutation relations of $T^I_{\ J}$,
$$ (-1)^{\hat K (\hat N+\hat L)}B^{IJ}_{\ \ MN}T^M_{\ K} T^N_{\ L}=
(-1)^{\hat M(\hat J+\hat N)}T^I_{\ M} T^J_{\ N} B^{MN}_{\ \ KL}, \eqno{(3.1)}$$
which was also discussed in ref.[18].

Using (2.20), we obtain explicit commutaion relations of $T^I_{\ J}$,
$$ T^I_{\ K} T^J_{\ K} =(-1)^{\hat K (\hat I +\hat J)}(-{1 \over X})^{\hat K}q
_{IJ}T^J_{\ K} T^I_{\ K},$$
$$ T^I_{\ K} T^I_{\ L} =(-1)^{\hat I (\hat K +\hat L)}(-{1 \over X})^{\hat I}
{ X \over q_{KL}}T^I_{\ L} T^I_{\ K}, $$
$$ T^I_{\ L} T^J_{\ K} =(-1)^{\hat K \hat I +\hat J \hat L}
{q_{IJ}q_{KL} \over X}T^J_{\ K} T^I_{\ L}, \eqno{(3.2)}$$
$$ T^I_{\ K} T^J_{\ L} =(-1)^{\hat K \hat J +\hat I \hat L}
{q_{IJ} \over q_{KL}}T^J_{\ L} T^I_{\ K} + (-1)^{\hat J(\hat K +\hat L)}
{X-1 \over q_{KL}}T^I_{\ L} T^J_{\ K},$$
where $I<J$ and $K<L$.
The relations show that $T^I_{\ J}$ are nothing but matrix elements of
$GL_q(m \vert n)$ quantum group, which has most deformation parameters
we have ever known.
It is remarkable that transposed matrix elements satisfy the same relations
except $q_{IJ}$ replacing $X/q_{IJ}$.

Next we define a superdeterminant of the quantum matrix $T$.
The superdeterminant for $GL_q(1|1)$ has been obtained in refs.[17,25].
Here we denote $T$ by block matrices,
$$ T^I_{\ J}= \left( \matrix { A^i_{\ j} & B^i_{\ \beta} \cr
C^\alpha_{\ j} & D^\alpha_{\ \beta} \cr}\right), \eqno{(3.3)}$$
where $A$ ($D$) transforms bosonic (fermionic) coordinates
into new bosonic (fermionic) coordinates.
At first we consider the case where only fermionic coordinates are linearly
transformed one another.
The fermionic determinant of the matrix $D$ is defined through an equation;
$$ \prod^n_{\alpha=1} \theta'^\alpha = {\rm det} _f D \prod^n _{\alpha =1}
\theta^\alpha.\eqno{(3.4)}$$
If we introduce a fermionic volume element ${\cal D}\theta$ as
$$ \int (\prod^n_{\alpha =1} \theta^\alpha) {\cal D}\theta=1,\eqno{(3.5)}$$
we find
$$ {\cal D}\theta'=({\rm det}_f D)^{-1} {\cal D} \theta.\eqno{(3.6)}$$
On the other hand, in transformation of only bosonic coordinates, the bosonic
determinant of the matrix $A$ is defined as ref.[11,12] through an equation,
$$ \prod^m_{k=1}dx'^k= {\rm det} _b A \prod^m_{k=1}dx^k .\eqno{(3.7)}$$
Then superdeterminant of the matrix $T$ is defined
using the above volume elements,
$$ (\prod^m_{k=1} dx'^k) {\cal D} \theta' = {\rm Sdet}T (\prod ^m_{k=1}dx^k)
{\cal D}\theta.\eqno{(3.8)}$$
Note that the matrix $T$ can be decomposeed into a product of three simple
matrices;
$$ \left( \matrix{ A & B \cr C & D \cr }\right) =
\left( \matrix{1 & BD^{-1} \cr 0 &1 \cr}\right) \left( \matrix{\tilde A & 0 \cr
 0 & D \cr}\right) \left( \matrix{1 & 0 \cr D^{-1}C & 1 \cr}\right) ,
\eqno{(3.9)}$$
where $\tilde A=A-BD^{-1}C$.
Only the second matrix of the right hand side contributes to
the superdeterminant, because the volume element is not changed
under the translation.
Thus we find Sdet$T=\det_ b \tilde A (\det _f D)^{-1}$.

Using (2.20), we obtain explicitly the bosonic and fermionic determinants as
$$ {\rm det} _b A =\sum_\sigma (\prod_{i<k,\sigma(i)>\sigma(k)}(-p_{\sigma(i)
\sigma(k)})^{-1})A^1_{\sigma(1)}A^2_{\sigma(2)}\cdots A^m_{\sigma(m)},
\eqno{(3.10.a)}$$
$$ {\rm det} _f D =\sum_\sigma
(\prod_{\alpha<\beta,\sigma(\alpha)>\sigma(\beta)}
(q_{\sigma(\alpha) \sigma(\beta)})^{-1})D^1_{\sigma(1)}D^2_{\sigma(2)}\cdots
D^n_{\sigma(n)},\eqno{(3.10.b)}$$
where $\sigma$ implies a possible permutation.
When $q_{IJ}=X=1$, the bosonic and the fermionic determinants
and the superdeterminant become \lq\lq classical" obviously.

Further,we obtain a commutaton relation,
$$ x^i \prod^m_{k=1}dx^k=(\prod ^m_{k=1}dx^k) x^i c_{(i)}X^i,\quad
d\theta^\alpha \prod^n_{\beta=1}\theta^\beta=(\prod^n_{\beta=1}\theta^\beta)
d\theta^\alpha c_{(\alpha)}X^{\alpha-n},\eqno{(3.11)}$$
where
$$ c_{(i)} \equiv  (\prod ^{i-1}_{k=1} {1 \over q_{ki}})\prod ^m _{k=i+1}
q_{ik}, \qquad c_{(\alpha)} \equiv  (\prod ^{\alpha-1}_{\beta=1}
{1 \over q_{\beta \alpha}})\prod ^n _{\beta =\alpha +1} q_{\alpha \beta}.
\eqno{(3.12)}$$
This leads to commutation relations between the bosonic determinant
and each matrix element $T^i_{\ j}$,
$$ c_{(j)} X^jT^i_{\ j} \ \det _b \tilde A=c_{(i)} X^i\ \det _b \tilde A
\ T^i_{\ j} ,\eqno{(3.13)}$$
which coincides with the result of ref.[12].
In addition to (3.11), we use the commutation relations of the transposed
matrix to find commutaion relations between the superdeterminant
and the matrix elements $T^I_{\ J}$.
Thus, we obtain relations,
$$ T^i_{\ j} {\cal T} = X^{i-j}{c_{(i)} \over c_{(j)}}(\prod ^n_{\beta=1}
{q_{j\beta} \over q_{i \beta}}) {\cal T} \ T^i_{\ j}, \quad T^\alpha_{\ \beta}
{\cal T} = X^{\beta-\alpha}{c_{(\beta)} \over c_{(\alpha)}}(\prod^m_{k=1}
{q_{k \beta} \over q_{k\alpha}}) {\cal T} T^\alpha_{\ \beta} ,$$
$$ T^i_{\ \alpha} {\cal T} = X^{i+\alpha-m-1}c_{(i)}c_{(\alpha)}{\prod^m_{k=1}
q_{k \alpha} \over \prod^n_{\beta=1} q_{i\beta}}{\cal T} T^i_{\ \alpha},
\eqno{(3.14)}$$
where ${\cal T} \equiv {\rm Sdet}T$.
We can find commutation relations between the superdeterminant and
$T^\alpha_{\ i}$, using the fact that the superdeterminant and
$T^I_{\ J} T^J_{\ I}$ commute each other.

The centrality of the superdeterminant requires
$${X^ic_{(i)} \over \prod^n_{\beta=1} q_{i \beta}}=
{X^{-\alpha-m-1}c_{(\alpha)}^{-1} \over \prod^m_{k=1} q_{k \alpha}}={\rm const}
,\eqno(3.15)$$
for all $i$ and $\alpha$.
In general, eq. (3.15) provides $(m+n-1)$ independent conditions
on the deformation parameters, while in special case where $m=n$ we find only
$(m+n-2)$ independent conditions.
Therefore, we can obtain $SL_q(m|n)$ quantum groups with $\{(N-2)(N-1)/2+1\}$
independent deformation parameters and $SL_q(m|m)$ quantum groups with
$\{(m-1)(2m-1)+2\}$ independent parameters.
The latter differs from the situation of $SL_q(m)$ of ref.[12] and that is
one of remarkable aspects of the quantum supergroups.

\vspace{0.8 cm}
\leftline{\large \bf 4. Deformed phase space}
\vspace{0.8 cm}

In this section, we deform phase space of the supersymmetric coordinates
and momenta, following refs.[10,19].
We consider the real quantum superspace, i.e., $\bar Z^I=Z^I$.
That requires conditions on the deformation parameters, i.e.,
$ \bar q_{IJ}=1/q_{IJ}$, $\bar X =1/X$.
Further, the derivatives should satisfy
$$ \bar \partial_i =-X^{2\rho(i)}\partial_i, \qquad \bar \partial_\alpha=
X^{2\rho(\alpha)} \partial_\alpha,\eqno{(4.1)}$$
where $\rho(i)=(m+1-n-i)/2$, $\rho(\alpha)=(\alpha-n)/2$
in the case of Type I ordering and $\rho(i)=(m+1-i)/2$,
$\rho(\alpha)=(m+\alpha-n)/2$ for Type II ordering.
Therefore we can define real momentum operators as
$$ p_i \equiv -i X^{\rho(i)}\partial_i, \qquad p_\alpha \equiv
X^{\rho(\alpha)}\partial_\alpha .\eqno{(4.2)}$$
Eqs.(2.2), (2.6) and (2.16) requires that these super phase-space should
follow commutation relations,
$$Z^IZ^J=B^{IJ}_{\ \ KL}Z^KZ^L,\qquad P_LP_K=B^{IJ}_{\ \ KL}P_JP_I,
\eqno{(4.3)}$$
$$P_KZ^I=-(i)^{\hat K+1} C^I_{\ K}+(i)^{\hat K-\hat J}
X^{\rho (K)-\rho(J)+1}B^{IJ}_{\ \ KL}Z^LP_J,$$
where $C^I_{\ K}=X^{\rho (K)}\delta^I_{\ K}$ and $P_I=(p_i, p_\alpha)$.

Suppose that we define super-gamma matrices as
$$ \gamma^\alpha=p_\alpha,\qquad \gamma^{2n-\alpha+1}=\theta^\alpha,\qquad c^i
=p_i,\qquad c^{2m-i+1}=x^i.\eqno{(4.4)}$$
Hereafter we denote $\alpha'=2n-\alpha+1$ and $i'=2m-i+1$.
We can derive deformed super-Heisenberg algebra or \lq\lq super"-Clifford
algebra from the phase space algebra as follows,
$$ \Gamma^I\Gamma^J+(-\tilde B^{IJ}_{\ \ KL})\Gamma^K\Gamma^L=
C^{IJ}(-i)^{\hat I+1},\eqno{(4.5)}$$
where $\Gamma^I=(\gamma^\alpha,c^i)$
$(\alpha=1, \cdots ,2n,\ i=1, \cdots , 2m)$.
The matrix $\tilde B^{IJ}_{\ \ KL}$ is ,of course, related to the matrix
$B^{IJ}_{\ \ KL}$, i.e.,
$$ \tilde B^{I'J'}_{\ \ K'L'}=B^{IJ}_{\ \ KL}, \qquad
\tilde B^{IJ}_{\ \ KL}=B^{LK}_{\ \ JI},\eqno(4.6.a)$$
$$ \tilde B^{IJ'}_{\ \ K'L}=(i)^{\hat I-\hat L} X^{\rho (I)-\rho(L)+1}
B^{JL}_{\ \ IK},\eqno{(4.6.b)}$$
$$ \tilde B^{I'J}_{\ \ KL'}=(i)^{\hat J-\hat K} X^{\rho (J)-\rho(K)-1}
(B^{-1})^{IK}_{\ \ JL},\eqno(4.6.c)$$
where $I,J,K,L<m+1$ ($n+1$) in the bosonic (fermionic) case and the inverse
of the matrix $B$ is defined as
$(B^{-1})^{LI}_{\ \ JK}B^{IM}_{\ \ KN}=\delta^{LM}\delta_{JN}$.
Further the metric matrix $C$ is obtained as $C^{IJ'}=C^J_{\ I}$, $C^{IJ}=0$
and so on.
Among the block matrices of the matrix $\tilde B$,
$\tilde B^{I'J'}_{\ \ K'L'}$ and $\tilde B^{IJ}_{\ \ KL}$ of (4.6.a) have
eigenvalues 1 and $-1/X$, as said in section two.
The other block matrix with (4.6.b) and (4.6.c) as off-diagonal sub-matrices,
has eigenvalues $\pm 1$.
It is easily shown that the matrix $\tilde B$ satisly the Yang-Baxter equation.
Suppose that we graphically represent the $\tilde B^{IJ'}_{\ \ K'L}$ of (4.6.b)
as Fig.2.
It is related to the diagram of $B^{JL}_{\ \ IK}$.
For example one set of the Yang-Baxter equations with indices shown in Fig.3
is identified with the equation represented by Fig.1.b,
up to an overall factor, $(i)^{\hat M-\hat N}X^{\rho (M)-\rho (N)+1}$.
Similarly, we can prove that the matrix $\tilde B$ with the other indices
satisfy the Yang-Baxter equation, making use of the fact that $B$ is
the solution of the Yang-Baxter equation and one sort of conservation law that
the matrix elements $B^{IJ}_{\ \ K L}$ vanish unless $I+J=K+L$ and $I-J=K-L$.
Imposing conditions,$q_{IJ}=q$ and $X=q^2$, we find that the $\tilde B$
matrices for the $GL_q(m|0)$ and $GL_q(0|n)$ phase spaces coincide
with well-known $\hat R$-matrices of $Sp_q(2m)$ and $SO_q(2n)$ deformed
by one parameter, respectively, as shown in ref.[19].
Further, the above analysis suggests $\hat R$-matrices for deformed
$OSp(2n|2m)$ groups.

Moreover, we can define a quantum superspace with a metric,
whose differentials satisfy commutation relations,
$$ d\tilde Z^I d\tilde Z^J =\tilde B^{IJ}_{\ \ KL} d\tilde Z^K d\tilde Z^L,
\eqno(4.7)$$
where new coordinates $\tilde Z^I$ in a deformed superspace consist of new
bosonic coordinates $\tilde x^\alpha$ $(\alpha=1 \cdots 2n)$ and new fermionic
coordinates $\tilde \theta ^i$ $(i=1 \cdots 2m)$, i.e., grassmann parities
of all new coordinates are reversed, compared with the orginal $GL_q(m|n)$
phase space.
{}From (4.7), we can derive commutation relations between the new coordinates
and the differentials by the discussion on (2.3) and (2.18) as
$$ \tilde Z^I d\tilde Z^J=-(-1)^{\hat K} \tilde B^{IJ}_{\ \ KL} d\tilde Z^K
\tilde Z^L. \eqno{(4.8)}$$
Further, using characteristic equations of the block matrices of $\tilde B$
and (2.11), we can derive commutaions relations for the new coordinates
$\tilde Z$ in a similar way to the procedure in the section two.

Let us consider the $GL_{q}(1|1)$ phase space as a concrete example,
from which we derive $\hat R$-matrix for deformed $OSp(2|2)$.
The deformation of $OSp(2|2)$ was also discussed algebraically in ref.[23].
In the case of Type I ordering, bosonic and fermionic momenta are defined as,
$$ p_x\equiv -i \partial_x,\qquad p_\theta \equiv \partial_\theta.
\eqno{(4.9)}$$
They and the coordinates satisfy commutaion relations as follows,
$$ p_x p_\theta ={q \over X}p_\theta p_x, \qquad x \theta = q \theta X,$$
$$p_x \theta = {X \over q} \theta p_x,\qquad p_\theta x= q x p_\theta,
\eqno(4.10)$$
$$p_x x=-i +X xp_x-i(X-1)\theta p_\theta,$$
$$p_\theta \theta =1- \theta p_\theta,\qquad p_\theta^2=\theta^2=0,$$
where these relations are dominated by the matrix $\tilde B$ constructed
as (4.6).
In this case, the matrix $\tilde B$ is obtained explicitly as
$$\tilde B=\pmatrix{a^* &       &   & \cr
                        & 0     & b & \cr
                        & b^{-1}& 0 & \cr
                        &       &   & a \cr},\eqno(4.11)$$
$$a=\pmatrix{-1/X & & & \cr & 0 & 1/q & \cr & q/X & 1-1/X & \cr & & & 1},
b=\pmatrix{-1 & & & \cr & 0 & q & \cr & X/q & 0& \cr i(1-X) & & & X \cr},$$
where $a^*$ is obtained from $a$ by rearrangement of the elements as (4.6.a).
Suppose that we define gamma matrices $\gamma^1 \equiv p_\theta$, $\gamma^2
\equiv \theta$, $c^1 \equiv p_x$ and $c^2 \equiv x$.
They satisfy super-Clifford algebra for deformed $OSp(2|2)$ group as
$$(\gamma^1)^2=(\gamma^2)^2=0,\qquad \gamma^1 \gamma^2 +\gamma^2 \gamma^1=1,$$
$$c^1c^2-Xc^2c^1+i(X-1)\gamma^2\gamma^1=-i,\eqno(4.12)$$
$$ c^1\gamma^1-{q \over X}\gamma^1c^1=0,\qquad \gamma^1 c^2 -q c^2 \gamma^1=0,
$$
$$ c^1 \gamma^2-{X \over q}\gamma^2 c^1=0, \qquad c^2 \gamma^2-q\gamma^2c^2=0.
$$
Using the matrix $\tilde B$ as (4.7), we can define commutation relations
between differentials of the deformed $OSp(2|2)$ superspace as,
$$(d\tilde x^1)^2=(d\tilde x^2)^2=0,\qquad d\tilde x^1 d\tilde x^2=
-d\tilde x^2 d\tilde x^1,$$
$$d\tilde \theta^1 d\tilde \theta^2=X d\tilde \theta^2 d\tilde \theta^1-i(X-1)
d\tilde x^2 d\tilde x^1,\eqno(4.13)$$
$$ d\tilde \theta^1 d\tilde x^1 = {q \over X} d\tilde x^1 d\tilde \theta^1,
\qquad  d\tilde \theta^1 d\tilde x^2 = {X \over q} d\tilde x^2
d\tilde \theta ^1,$$
$$ d\tilde x^1 d\tilde \theta^2=q d\tilde \theta^2 d\tilde x^1,
\qquad d\tilde x^2 d\tilde \theta^2 =q d\tilde \theta^2 d\tilde x^2.$$
Here we define another parity $\epsilon (I)$ as $\epsilon (1)=0$
and $\epsilon (2)=1$ in both bosonic and fermionic cases.
The matrix $\tilde B^{IJ}_{\ \ KL}$ satisfy the characteristic equation,
$$ (\tilde B^{IJ}_{\ \ KL}-\delta^I_{\ K} \delta^J_{\ K})
(\tilde B^{IJ}_{\ \ KL}+({1 \over X})^{|\epsilon (I)-\epsilon (J)|}
\delta^I_{\ K} \delta^J_{\ L})=0, \eqno{(4.14)}$$
and also $(-1)^{\hat I+\hat K}\tilde B^{IJ}_{\ \ KL}$ satisfy
the same equation as (4.14), where $\hat I$ and $\hat K$ denote
the grassmann parities of the new coordinates.
Using the characteristic equation (4.14), we can easily decompose the matrix
$\tilde B^{IJ}_{\ \ KL}$ into projection operators, as said in section two.
Then we will derive a matrix $\tilde B'^{IJ}_{\ \ KL}$ which dominates
commutation relations between new coordinates $\tilde Z^I$ as
$\tilde Z^I \tilde Z^J=\tilde B'^{IJ}_{\ \ KL} \tilde Z^K \tilde Z^L$.
Comparing the characteristic equation (4.14) with (2.11), we find
$$ \tilde B'^{IJ}_{\ \ KL}=-(-1)^{\hat I+\hat K}
X^{|\epsilon (I) -\epsilon (J)|}\tilde B^{IJ}_{\ \ KL}, \eqno{(4.15)}$$
and explicitly obtain the commutation relations as
$$\tilde x^1 \tilde x^2=\tilde x^2 \tilde x^1,\qquad (\tilde \theta^1)^2 =
(\tilde \theta^2)^2=0,$$
$$ \tilde \theta^1 \tilde x^1=q \tilde x^1 \tilde \theta^1, \qquad
\tilde x^1 \tilde \theta^2 =q \tilde \theta^2 \tilde x^1,\eqno(4.16)$$
$$ \tilde \theta^1  \tilde x^2 ={X \over q} \tilde x^2 \tilde \theta^1,
\qquad   \tilde x^2 \tilde \theta^2 ={X \over q} \tilde \theta^2 \tilde x^2,$$
$$  \tilde \theta^1 \tilde \theta^2=-X \tilde \theta^2 \tilde \theta^1-i(X-1)
\tilde x^1 \tilde x^2.$$
This algebra has a center,
$ \tilde x^1 \tilde x^2+i \tilde \theta^1 \tilde \theta^2$, i.e.,
it is not deformed from the classical one.
It can be described as
$$ \tilde x^i \tilde C_{ij} \tilde x^j+i \tilde \theta^\alpha
\tilde C'_{\alpha \beta} \tilde \theta^\beta \equiv {X+1 \over 4}
(\tilde x^1 \tilde x^2+\tilde x^2 \tilde x^1)+i{\sqrt X \over 2}
({1 \over \sqrt X} \tilde \theta^1 \tilde \theta ^2-\sqrt X \tilde \theta^2
\tilde \theta^1). \eqno(4.17)$$
It is remarkable that ratios of the metrics, $\tilde C_{12}/ \tilde C_{21}$
and $\tilde C'_{12}/ \tilde C'_{21}$, coincide with those of $SO_q(2)$ and
$Sp_q(2)$ quantum groups of ref.[7], respectively.
By similar procedure, we can derive commutaion relations of the deformed
$OSp(2|2)$ superspace for Type II ordering, where we can find that
$\tilde x^1 \tilde x^2+i\tilde \theta^1 \tilde \theta^2$ is also central.
Further the similar analysis on the $GL_{q}(m|n)$ phase space lead to
various quantum superspaces with metrics.

\vspace{0.8 cm}
\leftline{\large \bf 5. Concluding remarks}
\vspace{0.8 cm}

  In this paper we investigated the differential calculus on the
non-commutative superspace. We obtained the Yang-Baxter equation
in this framework for quantum supergroups, and in particular, the explicit
form for the multiparametric ${\hat R}$-matrix in the case of q-deformed
general linear supergroup, $GL_q(m|n)$. We also presented the commutation
relation of quantum supergroup matrix-elements as well as quantum
superdeterminants.  We then extended these results to phase-space construction
of $OSp_q(2n|2m)$. Some details for quantum super-Clifford algebras and
the ${\hat R}$-matrix were presented for the case of $OSp_q(2|2)$.

  It would be interesting to extend the present result to more general
supergroups.
It should also be worked out to find the relation between the quantum group
and the quantum Lie algebras. In the context of differential calculus, one of
the next issues to be studied is the quantum differential form
\cite{Zumino2}.

  Finally, there have been a number of works on quantum
deformations of space-time symmetries such as Lorentz
\cite{Carow-Schlieker-Scholl-Watamura,Podles-Woronowicz,Schmidke-Wess-Zumino}
and Poincar{\' e}
\cite{Lukierski-Ruegg-Nowicki-Tolstoy,Lukierski-Nowicki-Ruegg,OSWZ} groups
in the literatures.
In this connection, the most intriguing subject to be pursued further might
be q-deformation of super-Poincar{\' e} symmetry.

\vspace{0.8 cm}

 The authors would like to thank T. Inami, P. Kulish and R. Sasaki for
valuable discussions.

\newpage
\hoffset = 1 cm
\begin{figure}[hbt]
\setlength{\unitlength}{1mm}
\begin{picture}(80,20)(-15,-5)
\thicklines
\put(5,-5){\line(1,-1){10}}
\put(5,-15){\line(1,1){4}}
\put(11,-9){\line(1,1){4}}
\bf
\put(2,-5){I}
\put(15,-5){J}
\put(1,-16){K}
\put(15,-16){L}
\put(-20,-10){B$^{\bf IJ}_{\ \ \bf KL} \ = \ $}
\put(10,-29){a}
\put(-25,-35){Fig.1.a,b \  {\rm Graphical representations of the matrix $B$
and the Yang-Baxter equation}}
\end{picture}
\setlength{\unitlength}{1mm}

\begin{picture}(100,20)(-65,-5)
\thicklines
\put(9.2,9.2){\line(-1,-1){3.6}}
\put(0,20){\line(1,-1){20}}
\put(5,20){\line(0,-1){4}}
\put(5,14){\line(0,-1){14}}
\put(0,0){\line(1,1){4}}
\put(11,11){\line(1,1){9}}
\put(5,10){\circle*{1}}
\put(7.5,7.5){\circle*{1}}
\put(7.5,12.5){\circle*{1}}
\put(0,9){\bf K}
\put(8,4){\bf S}
\put(8,14){\bf L}
\put(-3,20){\bf I}
\put(8,20){\bf J}
\put(21,20){\bf N}
\put(-3,-3){\bf M}
\put(8,-3){\bf Q}
\put(21,-3){\bf V}
\put(30,10){$=$}
\put(30,-9){\bf b}
\put(40,20){\line(1,-1){20}}
\put(55,0){\line(0,1){4}}
\put(55,6){\line(0,1){14}}
\put(40,0){\line(1,1){9}}
\put(50.8,10.8){\line(1,1){3.6}}
\put(56,16){\line(1,1){4}}
\put(55,10){\circle*{1}}
\put(52.5,7.5){\circle*{1}}
\put(52.5,12.5){\circle*{1}}
\put(37,20){\bf I}
\put(50,20){\bf J}
\put(61,20){\bf N}
\put(37,-3){\bf M}
\put(50,-3){\bf Q}
\put(61,-3){\bf V}
\put(50,14){\bf P}
\put(50,4){\bf T}
\put(57,9){\bf U}

\end{picture}
\end{figure}
\setlength{\unitlength}{1mm}
\begin{picture}(80,20)(-15,-5)
\thicklines
\put(5,-5){\line(1,-1){10}}
\put(5,-15){\line(1,1){4}}
\put(11,-9){\line(1,1){4}}
\bf
\put(1,-5){I}
\put(15,-5){J$'$}
\put(0,-16){K$'$}
\put(15,-16){L}
\put(-22,-10){$\tilde {\bf B}^{{\bf IJ}'}_{\ \ {\bf K}'{\bf L}} \ = \ $}
\put(10,-10){\circle{3}}
\put(25,-10){$ = {\bf (i)}^{\hat {\bf I} - \hat {\bf L}}
{\bf X}^{\rho ({\bf I})-\rho({\bf L})+1}$}
\put(75,-5){\line(1,-1){10}}
\put(75,-15){\line(1,1){4}}
\put(81,-9){\line(1,1){4}}
\put(71,-5){J}
\put(85,-5){L}
\put(72,-16){I}
\put(85,-16){K}

\put(-25,-35){Fig.2 \  {\rm Graphical representation of the block matrix of
$\tilde{\bf B}$ written as eq(4.6.b)}}
\end{picture}
\setlength{\unitlength}{1mm}

\begin{picture}(100,20)(0,40)
\thicklines
\put(9.2,9.2){\line(-1,-1){3.6}}
\put(0,20){\line(1,-1){20}}
\put(5,20){\line(0,-1){4}}
\put(5,14){\line(0,-1){14}}
\put(0,0){\line(1,1){4}}
\put(11,11){\line(1,1){9}}
\put(5,15){\circle{3}}
\put(10,10){\circle{3}}
\put(-3,20){\bf M}
\put(8,20){\bf I$'$}
\put(21,20){\bf J$'$}
\put(-3,-3){\bf Q$'$}
\put(8,-3){\bf V$'$}
\put(21,-3){\bf N}
\put(30,10){$=$}
\put(40,20){\line(1,-1){20}}
\put(55,0){\line(0,1){4}}
\put(55,6){\line(0,1){14}}
\put(40,0){\line(1,1){9}}
\put(50.8,10.8){\line(1,1){3.6}}
\put(56,16){\line(1,1){4}}
\put(55,5){\circle{3}}
\put(50,10){\circle{3}}
\put(37,20){\bf M}
\put(50,20){\bf I$'$}
\put(61,20){\bf J$'$}
\put(37,-3){\bf Q$'$}
\put(50,-3){\bf V$'$}
\put(61,-3){\bf N}
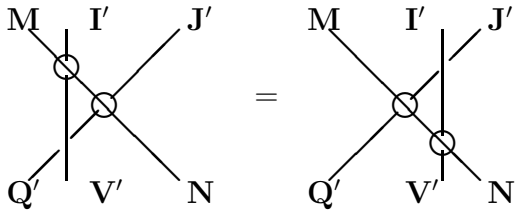
\put(-15,-10){\bf Fig.3 \ {\rm Graphical representation of some of the
Yang-Baxter equations of $\tilde B$}}

\end{picture}


\end{document}